\documentclass[11pt]{article}
\usepackage{amsmath,moriond,epsfig}

\begin{document}
\vspace*{4cm}

\title{Charmless Two-body $B(B_s)\to PP,VP$ decays In Soft-Collinear-Effective-Theory}
\author{ Wei Wang$^{a,b}$, Yu-Ming Wang$^{a,b}$, De-Shan Yang$^b$ and Cai-Dian L\"u$^a$ }
\address{
 $^a$ Institute of High Energy Physics, Chinese Academy
of Sciences, Beijing 100049, P.R. China\\
 $^b$ Graduate University of Chinese Academy of
Sciences, Beijing 100049, P.R. China}

\maketitle

\abstracts{ We   analyze the charmless two-body $B\to VP$ decays in
the soft-collinear-effective-theory (SCET), where $V(P)$ denotes a
light vector (pseudoscalar) meson.
  Using the current
  experimental data, we find two
solutions in $\chi^2$ fit for the 16 non-perturbative inputs
responsible for the 87 $B\to PP$ and $B\to VP$ decay channels in
SU(3) symmetry. Chirally enhanced penguins can only change several
charming penguins sizably, since they share the same topology. The
$(S-P)(S+P)$ annihilation penguins in
    the perturbative QCD approach have the same topology with charming penguins
 in SCET, which play an important role in direct CP asymmetries.
 }

\section{Introduction}

The charmless two-body non-leptonic $B$ decays are important for the
precise test of the standard model   and the search for possible new
physics signals.   To predict branching ratios and CP asymmetries,
one has to compute the hadronic decay amplitudes. In recent years,
great progresses have been made in studies of charmless two-body $B$
decays, such as the generalized factorization
approach,\cite{Ali:1997nh} the QCD factorization
(QCDF),\cite{Beneke:1999br,Beneke:2003zv} the perturbative QCD
(pQCD) \cite{Keum:2000ph} and the soft-collinear effective theory
(SCET).\cite{Bauer:2000yr} Despite of many differences, all of them
are based on power expansions in $\Lambda_{QCD}/m_b$. Factorization
of the hadronic matrix elements is proved to hold in the leading
power in $\Lambda_{QCD}/m_b$ in a number of decays.

  It is almost an impossible task to
include all power corrections, but we can include the relatively
important one. Importance of chirally enhanced penguins has been
noted long time ago, and numerics show that chiraly enhanced
penguins are comparable with the penguin contributions at leading
power, in both of QCDF~\cite{Beneke:1999br,Beneke:2003zv} and
pQCD~\cite{Keum:2000ph} approaches. In SCET, the complete operator
basis and the corresponding factorization formulae for this term are
recently derived.\cite{Arnesen:2006vb} A new factorization formula
for chiraly enhanced penguin was proved to hold to all orders in
$\alpha_s$, and more importantly the factorization formula does not
suffer from the endpoint divergence.

One phenomenological framework is introduced not using the expansion
at the intermediate scale $\mu_{hc}=\sqrt
{m_b\Lambda_{QCD}}$.\cite{Bauer:2005kd} Instead the experimental
data are used to fit the non-perturbative inputs. This method is
very useful especially at tree level,  where only a few inputs are
required in decay amplitudes. In this framework, an additional term
from the intermediate charm quark loops, which is called charming
penguin,\cite{Bauer:2005kd,Bauer:2004tj,Bauer:2005wb} is also taken
into account. Charming penguins are not factorized into the LCDAs
and form factors, since the heavy charm quark pair can not be viewed
as collinear quarks. They are also treated as non-perturbative
inputs.  This method is first applied to $B\to K\pi$, $B\to KK$ and
$B\to\pi\pi$ decays,\cite{Bauer:2005kd} and later   to charmless
two-body $B\to PP$ decays involving the iso-singlet mesons $\eta$
and $\eta'$.\cite{Williamson:2006hb}

In the present work, we extend this method to the $B\to VP$ decays,
including the chirally enhanced penguins. We will use the wealth of
the experimental data to fit the non-perturbative inputs (in our
analysis, we also take the $B\to PP$ decays into
account).\cite{wang}

\section{$B\to VP$ decay amplitudes at leading power in
SCET}\label{sec:decayamplitudes}

  The weak effective Hamiltonian which
describes $b\to D$ ($D=d,s$) transitions is:\cite{Buchalla:1995vs}
 \begin{eqnarray}
 {\cal H}_{eff} &=& \frac{G_{F}}{\sqrt{2}}
     \bigg\{ \sum\limits_{q=u,c} V_{qb} V_{qD}^{*} \big[
     C_{1}  O^{q}_{1}
  +  C_{2}  O^{q}_{2}\Big]- V_{tb} V_{tD}^{*} \big[{\sum\limits_{i=3}^{10,7\gamma,8g}} C_{i}  O_{i} \Big]\bigg\}+ \mbox{H.c.} ,
 \label{eq:hamiltonian}
\end{eqnarray}
where $V_{qb(D)}$ are the CKM matrix elements.

  In $B\to M_1M_2$
decays, both of the final state mesons move very fast and are
generated back-to-back in the rest frame of $B$ meson.
Correspondingly, there exist three typical scales: the $b$ quark
mass $m_b$, the soft scale $\Lambda_{QCD}$ set by the typical
momentum of the light degrees of freedom in the heavy $B$ meson, the
intermediate scale $\sqrt{m_b\Lambda_{QCD}}$ which arise from the
interaction between collinear particles and soft modes. SCET
provides an elegant theoretical tool to separate the physics at
different scales and factorization  for $B\to M_1M_2$ was proved to
hold to all orders in $\alpha_s$ at leading power of
$1/m_b$.\cite{Bauer:2004tj,Bauer:2001yt} After integrating out the
fluctuations with off-shellness $m_b^2$, one reaches the
intermediate effective theory SCET$_I$,   where the generic
factorization formula for $B\to M_1 M_2$   is written by:
\begin{eqnarray}
 \langle M_1 M_2|O_i|B\rangle= T(u) \otimes \phi_{M_1}(u)\zeta^{B\to M_2}
 +T_J(u,z) \otimes \phi_{M_1}(u)\otimes \zeta_J^{B\to M_2}
 (z),\label{eq:genericfactorization}
\end{eqnarray}
with $T$ and $T_J$ the perturbatively calculable Wilson
coefficients.   In the second step, the fluctuations with typical
off-shellness $m_b\Lambda_{QCD}$ are integrated out and one reaches
SCET$_{II}$. In SCET$_{II}$, end-point singularities prohibit the
factorization of $\zeta$, while the function $\zeta_J$ can be
further factorized into the convolution of a hard kernel (jet
function) with light-cone distribution amplitudes:
\begin{eqnarray}
 \zeta_J(z)=\phi_{M_2}(x)\otimes J (z,x,k_+)\otimes \phi_B(k_+).
\end{eqnarray}

\begin{figure}
%\rule{5cm}{0.2mm}\hfill\rule{5cm}{0.00mm}
\includegraphics[scale=0.45]{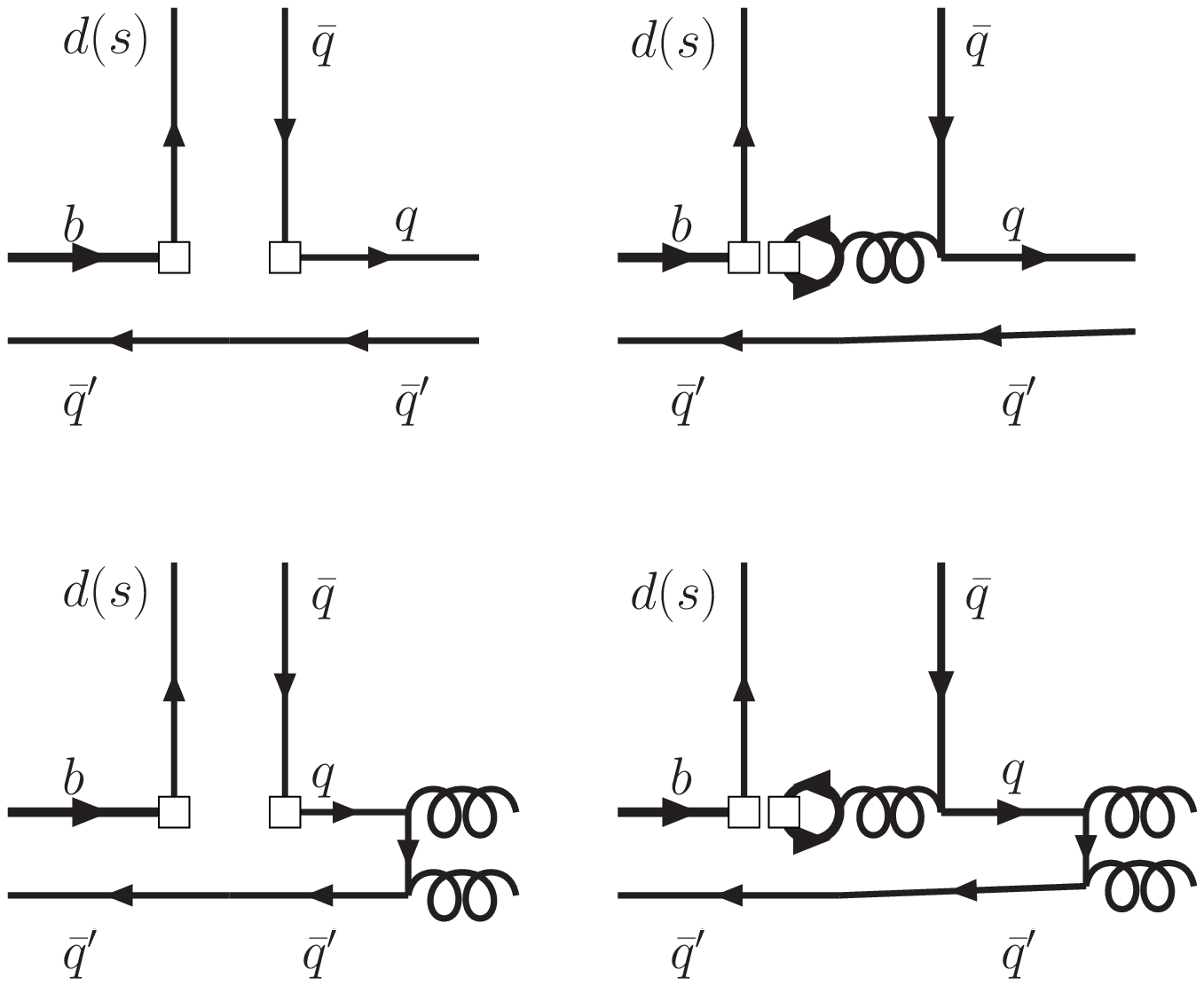}
\hspace{1cm}
\includegraphics[scale=0.25]{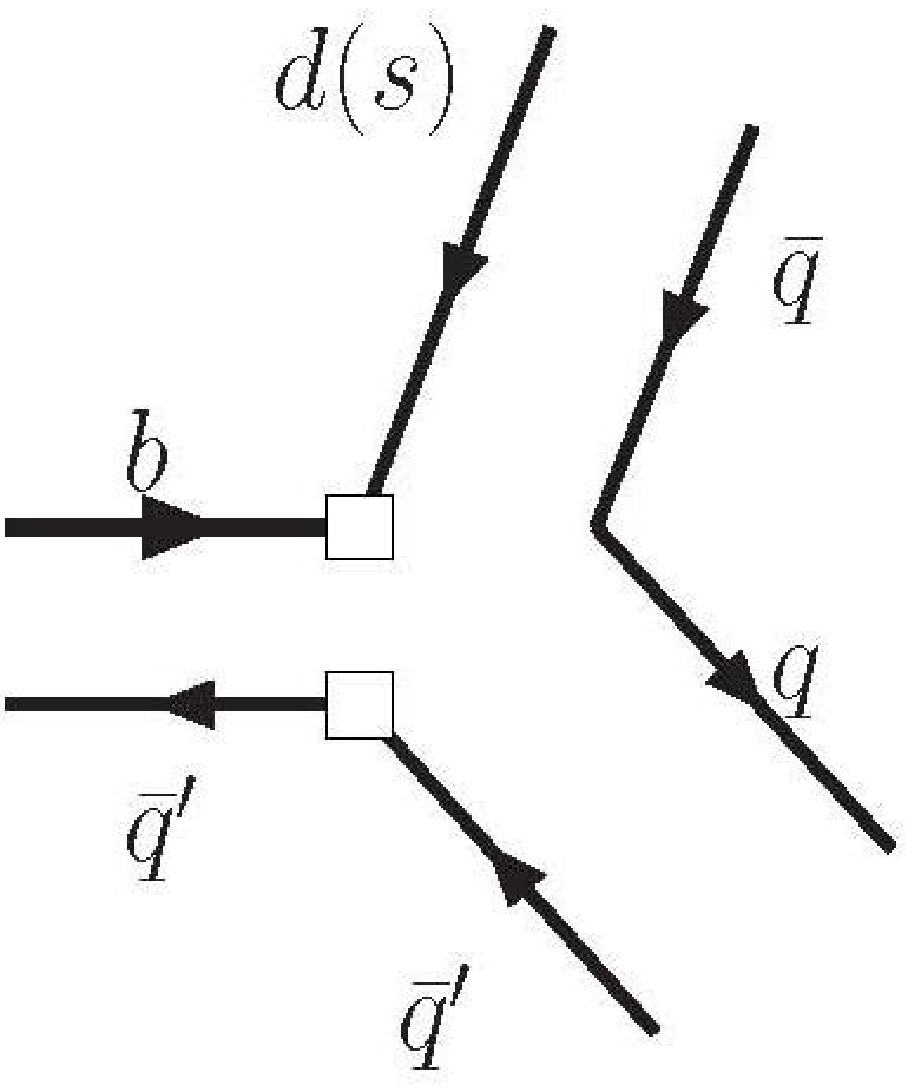}
%\rule{5cm}{0.00mm}\hfill\rule{5cm}{0.00mm}
%\psfig{figure=filename.ps,height=1.5in}
\caption{Feynman diagrams for chiraly enhanced penguins (left),
charming penguins (center) and the $(S-P)(S+P)$ annihilation
penguins (right)    } \label{diagram}
\end{figure}
%===============================================================

In order to reduce the independent inputs, one can utilize the SU(3)
symmetry for $B$ to light form factors and charming penguins.  In
the exact SU(3) limit, only two form factors are needed for $B\to
PP$ decays without iso-singlet mesons  $ \zeta_{(J)}^{BP} \equiv
\zeta_{(J)}^{B\pi}= \zeta_{(J)}^{BK}=\zeta_{(J)}^{B_sK}$. There are
two additional   non-perturbative functions $\zeta_{(J)g}$ in decays
involving iso-singlet mesons $\eta_q$ and $\eta_s$  from the
intrinsic gluon contributions. Since there is no gluonic
contribution in vector meson, there are only two $B\to V$ form
factors $ \zeta_{(J)}^{BV} \equiv\zeta^{B\rho}_{(J)}=\zeta^{B
K^*}_{(J)}=\zeta^{B\omega}_{(J)}
 =\zeta^{B_s  K^*}_{(J)}=\zeta^{B_s\phi}_{(J)}$
 in the SU(3) symmetry  and
  five complex charming penguins.
With the assumption of flavor SU(3) symmetry for $B$ to light form
factors and charming penguin terms, the non-perturbative, totally 16
real inputs responsible for $B\to PP$ and $B\to VP$ decays are
 $
   \zeta^{BP}$, $\zeta_J^{BP}$, $\zeta_g$, $\zeta_{Jg}$, $
    \zeta^{BV}$, $\zeta^{BV}_{J}$, $A^{PP}_{cc}$, $
    A^{PV}_{cc}$, $A^{VP}_{cc}$, $ A_{ccg}^{PP}$, $
    A^{VP}_{ccg}$.

Power corrections are expected to be suppressed by at least the
factor $\Lambda_{QCD}/m_b$, but chiraly enhanced penguins are large
enough to compete with the leading power QCD penguins as the
suppression factor becomes $2\mu_P/m_b$, where $\mu_P\sim 2$ GeV is
the chiral scale parameter. The complete operator basis and the
corresponding factorization formulae for the chiraly enhanced
penguin are recently derived.\cite{Arnesen:2006vb} The factorization
formula will introduce a new form factor $\zeta_{\chi}$ and a new
light-cone distribution amplitude $\phi^{pp}$.

\section{Numerical analysis of  $B\to VP$ decays}
\label{sec:numerical}

With experimental data for branching fractions and CP asymmetries,
$\chi^2$ fit method is used to determine the non-perturbative
inputs: form factors and charming penguins. Straightforwardly, we
obtain   two solutions for numerical results of the 16
non-perturbative inputs. As shown in Fig.~\ref{diagram}, chiraly
enhanced penguins have the same topology with the charming penguins.
The two diagrams in the lower line only contribute to decays
involving $\eta$ or $\eta'$, where $q=q'$. The inclusion of chirally
enhanced penguin will mainly change the size of three charming
penguins $A_{cc}^{PP}$, $A_{ccg}^{PP}$, $A_{cc}^{PV}$. Predictions
for branching fractions and CP asymmetries will not be changed
sizably.\cite{wang}

Our predictions for branching ratios of $\bar B^0\to\pi^0\rho^0$ are
larger than that in QCDF.  The tree contribution proportional to the
soft form factor $\zeta$ is color-suppressed, thus the branching
fractions of $\bar B^0\to\pi^0\rho^0$ in QCDF approach and pQCD
approach are much smaller than  ${\cal BR}(\bar
B^0\to\rho^\pm\pi^\mp)$. One important feature of the  SCET
framework is that the hard-scattering form factor $\zeta_J$ is
relatively large and comparable with the soft form factor $\zeta$.
Besides, this term has a large Wilson coefficient $b_1^f\sim 1.23$,
it can give larger production rates which are consistent with the
present experimental data. The agreement is very encouraging. We
also predict larger branching ratios for color-suppressed $B_s$
decays than QCDF \cite{Beneke:2003zv} and pQCD \cite{Ali:2007ff}
which can be tested on the future experiments.

For the decays with sizable branching fractions, our predictions on
direct CP asymmetries are typically small and most of them have the
correct sign with experimental data. Predictions in QCDF approach on
these channels are also small in magnitude, but most of them have
different signs with our results and experimental data. In pQCD
approach, the strong phases mainly come from the $(S-P)(S+P)$
annihilation operators. These operators are chiraly enhanced and the
imaginary part are dominant. Thus the direct CP asymmetries in pQCD
approach are typically large in
magnitude.\cite{Keum:2000ph,Ali:2007ff}

In pQCD approach, annihilation diagrams can be directly calculated.
The large $(S-P)(S+P)$ annihilation penguin operators   can explain
the correct branching ratios and direct CP asymmetries of
$B^0\to\pi^+\pi^-$ and $\bar B^0\to K^-\pi^+$,\cite{Hong:2005wj} the
polarization problem of $B\to \phi K^*$,\cite{Li:2004mp} etc. In
Fig.~\ref{diagram}, we draw the Feynman diagrams for this term.
Comparing with charming penguins, we can see they have the same
topologies in flavor space. Charming penguins in SCET as shown in
Fig.~\ref{diagram} play the similar role with $(S-P)(S+P)$
annihilation penguin operators in pQCD. But, the CKM matrix elements
associated with charming penguins and $(S-P)(S+P)$ annihilation
penguin operators are   proportional to $V_{tb}V_{tD}^*$ and
$V_{cb}V_{cD}^*$, respectively. The differences in the CKM matrix
elements will affect direct CP asymmetries and mixing-induced CP
asymmetries sizably, which can be tested at the future experiments.

\section{Conclusions}\label{sec:conclusions}

In the
 soft-collinear-effective theory,
we   analyze  the charmless two-body $B \to PP$, $B\to VP$ decays
 by taking some power corrections (chiraly
enhanced penguins) into account.    Using the  experimental data on
branching fractions and CP asymmetry variables, we find two
solutions in $\chi^2$ fit for the 16 non-perturbative inputs.
Chiraly enhanced penguin could change some charming penguins
sizably, since they have the same topology with each other. However,
most of other non-perturbative inputs and predictions on branching
ratios and CP asymmetries are not changed too much. With the two
sets of inputs, we predict branching fractions and CP asymmetries.
Agreements and differences with results in QCD factorization and
perturbative QCD approach are also analyzed.
 For example, we predict  larger branching ratios for
$B^0\to\pi^0\rho^0$ than QCDF and pQCD approach, but our reulsts are
consistent with the experimental data.

The $(S-P)(S+P)$ operators annihilation penguins  provide the main
strong phase in pQCD approach. In the SCET framework, charming
penguins play similar role especially the strong phase in $b\to s$
transitions. The $(S-P)(S+P)$ annihilations have the same topology
with charming penguin. There are alao differences in these two
objects including weak phases,  strong phases, SU(3) symmetry
property and factorization property. These differences will be
tested in the future experiments.

%%%%%%%%%%%%%%%%%%%%%%%%%%%%%%%%%%%%%%%%%%%%%%%%%%%%%%%%%%%%%%%
\section*{Acknowledgements}\vspace{-.2cm}
This work is partly supported by National Nature Science Foundation
of China under the Grant Numbers 10735080, 10625525 and 10705050.

\end{document}